# Latent Semantic Search and Information Extraction Architecture

Anton Kolonin[1]

[1]*Novosibirsk State University, 1 Pyrogova str., Novosibirsk, 630090, Russia.*

akolonin@gmail.com

**Annotation.** *The motivation, concept, design and implementation of latent semantic search for autonomous software agents with artificial intelligence is described. It is considered that mainstream search engines have limited semantic search, entity extraction and property attribution features, have insufficient accuracy and response time of latent search, may impose privacy concerns and the search results are unavailable in offline mode for robotic search operations. The alternative suggestion involves autonomous search engine with adaptive storage consumption, configurable search scope and latent search response time with built-in options for entity extraction and property attribution available as open source platform for mobile, desktop and server solutions. The suggested architecture attempts to implement artificial general intelligence (AGI) principles as long as autonomous behavior constrained by limited resources is concerned, and it is applied for specific task of enabling Web search for artificial agents implementing the AGI.*

## 1 Introduction

Since the origins of the information retrieval technology and establishment of the Web search as we know it today [1], the importance of information extraction was known as a necessary part [2]. That is, besides ability to find information of the interest, the abilities to classify it correctly, identify entities in the found information and determine properties of these entities correctly remains necessary part of the information processing [3] highly demanded on the mass market.

While many solution solving the problems mentioned above are known [4], not all of them provide sufficient accuracy. The less solutions provide capability of latent search [5] being able to monitor Web environment for target information constantly and providing updates upon appearance of novel data. Even with exiting Web alerting solutions, response times of latent search may be not sufficient for time-critical applications so that when the new information appears somewhere on the Web search engine may need hours to days in order to discover the change upon the next crawling cycle given the volume of world-wide volume indexing.

The other concern specific to solutions based on cloud services is related to security and privacy [6] so that some of the practical cases require the search requests and search results are sensitive to the searcher and may not be trusted to the cloud service provider.

From practical standpoint, the other useful feature missed by cloud-based Web search solutions is offline mode where it could be be possible to deal with search results and extracted information without of live connection to the cloud [7]. Also, major cloud-based search providers are very restrictive to search automation in general and making the search results available for robotic consumption so no any major search engine makes it officially possible for artificial intelligence software issue and process search requests and results.

Potentially, the solution of some of the latter problem would be implementation of the Web search solutions developed and/or hosted in-house, however straightforward reproduction of cloud-based search architectures [1] may cause inapplicable deployment and maintenance costs. From this perspective, adaptive search solutions for autonomous computing agents, capable to intelligent goal-driven resource allocation seem to be demanded in the short-term future of software agents powered by artificial intelligence (AI) [8].

As soon as adaptive goal-driven behavior is mentioned, it worth considering the criteria for posing the goals for the autonomous agents with AI to reach with the information retrieval and extraction operations. Here, the deep personalization personalization of the search operations may be enabled with enforced privacy and security specific to autonomous solutions hosted in-house or on privately owned equipment.

Finally, the deep personalization of the private and secure search for corporate systems or systems used by specific social groups may be enhanced with power of "social computing" where "crowd-sourcing" may turn the search system into recommendation system implicitly via algorithms like "page rank" [9] or explicitly my means of content relevance assessment by social values [10].

The following discussion will be dedicated to the concepts, design and implementation of the prototype Web search system capable to do semantic search, entity extraction and attribution [3] enforcing security and privacy of the users by means of custom hosting an any computational device, implementing principles of adaptive goal-driven information indexing, based on reinforcement learning from feedback provided by user our community of users with scope of social feedback turning the system into recommendation system.

## 2 Architecture and Implementation

The architecture of the adaptive latent semantic web search intended to address problems mentioned above is being implemented as part of Aigents project [8,10] since 2014 and it is available as open source since 2018 https://github.com/aigents/aigents-java.

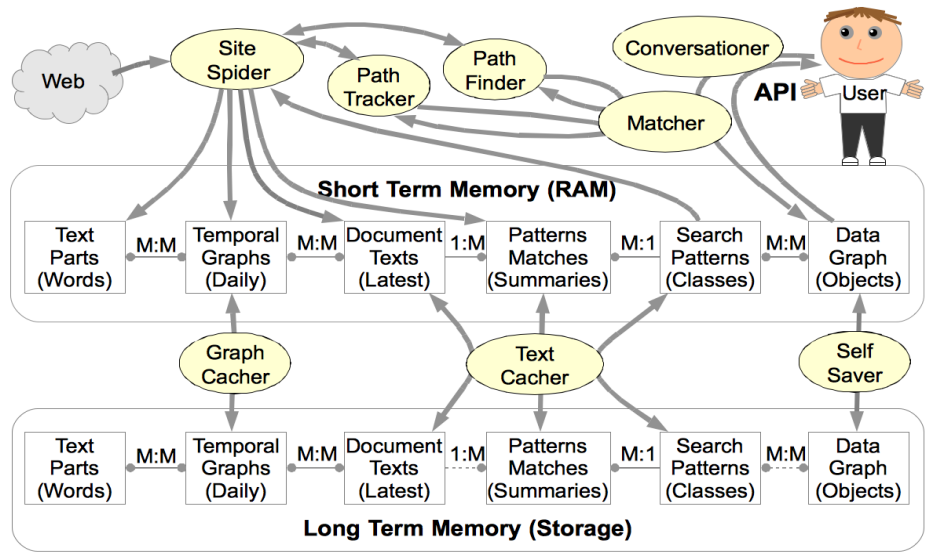

**Fig. 1.** Brief architecture of the adaptive latent semantic web search with STM and LTM layers and components serving Web crawling (Site Spider), adaptive search (Path Tracker and Path Finder), patter matching and semantic extraction (Matcher), interaction with user or integration API (Conversationer) as well as serialization and actualization of persistent data (Graph Cacher, Text Cacher and Self Saver).

The foundation of the architecture is temporal graph database used to store relationships between Web pages, pieces of text and semantic concepts and relationships associated into patterns and objects as shown on Fig. 1. The most volume of semantic data describing subject domain of the search application resides in Data Graph implemented as in-memory triplet store ins short-term memory (RAM) incrementally backed up on long-term persistent storage. Temporal history of web crawling results and extracted textual and semantic information is residing in dedicated Temporal Graph which is persistent in the long-term storage but can be partially loaded into RAM depending on attention focus of the system due to interactions with user or consumer application via API, ongoing latent search activity and available system resources. While short-term memory (STM) and long-term memory (LTM) in context of the search architecture are corresponding to RAM and persistent storage, on a broader scale of the architecture of the AI agent hosting the engine they may be corresponding to actual STM and LTM in terms of attention allocation and cognitive architecture [10].

The principal feature is latent targeted adaptive semantic web search on itself which can be broken in few parts, as follows.

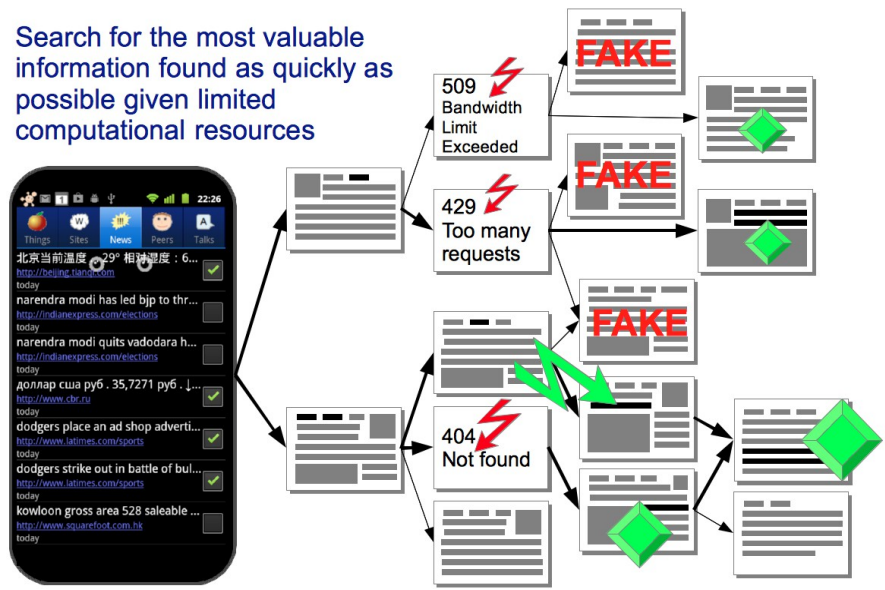

**Fig. 2.** Specialization of artificial generic intelligence problem to the Web search domain as "ability to reach target Web pages in complex Web environment with multiple links with different text descriptions and response time, given limited time and memory".

First, the latent operation is made possible with user or external API not just submitting search queries but having list of search queries or matching patterns to be configured along with potentially relevant starting URLs. These together identify the segment of the entire internet, accessible from the starting URLs and matching the configured queries and patterns. Moreover, the system spiders the internet using "targeted adaptive search" algorithm [8].

The **targeted adaptive search** on itself implements paradigm of artificial general intelligence (AGI) defined as "ability to reach complex goals in complex environments given limited resources", being grounded to subject domain of Web search and page surfing, as shown on Fig. 2. This is implemented with Path Finder capable to find Web surfing paths on the network of Web Links given the list of target queries and templates and Path Tracker capable to follow paths that are known as leading to targets [12]. Switching between strategies of Path Tracker, (shown as Algorithm 1 on Fig.3.) for known paths and Path Tracker (Algorithm 2, Fig. 3.) exploring new possible paths adaptively depends on available search time and may be also controlled by user and consumer applications.

Both strategies above start with goal pattern to match, starting Web page, and optional "path set" as collection of "paths". Each path is considered as chain of Web links that may be followed when surfing the web page with every link on the path identified by pattern on itself. Both strategies operate identifying Web page textual context as well a textual contexts of every Web link on the page matching them with goal patterns and patterns building the paths of the "path set". The strategies may be constrained by parameter of search modality defining whether the search should be "quick" to return the first matching result or "exhaustive" to get all possible results.

Additional constrains may be given by limit on search depth, search time and whether to follow only the Web links under the same Web domain as starting page or any domains can be explored.

Pattern matching is done with Matcher component following the earlier work [3] with generic hierarchical patterns which may consist of keywords, regular expressions, variables and ordered sets and variants of all of the above. In turn, variables may be restricted to belong to specific domains referenc-ing the other patterns and may correspond to semantic objects (entities) stored in graph database or properties of these objects.

```
Algorithm 1 PathTracker (recursive)
Input: Goal pattern, starting page, known "path set"
Parameter: modality ("exhaustive" or "quick")
Output: "result set" of pages with pattern matches
1:  Get the "page context" from page
2:  Evaluate "page context" against the goal pattern
3:  If "page context" is matching the goal:
4:      Add matching results to "result set"
5:      If "modality" is not "exhaustive":
6:          Return "result set"
7:  Get "path set" leading to goal
8:  Get all "link contexts" from page
9:  Evaluate "path set" items against the "link contexts"
    from the shortest path in "path set" to longest one:
10:     For each "link context" matching "path set" item:
11:         Reduce "path set" excluding the item ("burnout")
12:         Follow the link of "link context" to new page
13:         Run Path Tracker recursion (same goal pattern,
            new page, new "path set")
14:         On successful return:
15:             If modality is not "exhaustive":
16:                 break
17: If "result set" is not empty:
18:     Return "result set"
19: If not recursing:
20:     Run PathFinder
```

```
Algorithm 2 Path Finder (recursive)
Input: Goal pattern, starting page, hypothetical "path set"
Parameter: modality ("exhaustive" or "quick")
Output: "result set" of pages with pattern matches and new "path set"
1:  Get the "page context" from page
2:  Evaluate "page context" against the goal pattern
3:  If "page context" is matching the goal:
4:      Add matching results to "result set"
5:      Retain the hypothetical "path set" as proven
6:      If "modality" is not "exhaustive":
7:          Return "result set" and retained "path sets"
8:  Get all "link contexts" from page
9:  Evaluate all possibilities extending hypothetical
    "path set" with new "link contexts":
10:     For each "link context":
11:         Add new "link context" to current "path set"
            creating hypothetical "path set"
12:         Follow the link of "link context" to new page
13:         Run Path Finder recursion (same goal pattern,
            new page, new hypothetical "path set")
14:         On successful return:
15:             If modality is not "exhaustive":
16:                 break
17: If "result set" is not empty:
18:     Return "result set" and retained "path set"
19: If not recursing:
20:     Merge retained "path sets" with original one known
        for the goal
```

**Fig. 3.** Explorative "PathFinder" and conservative "PathTracker" algorithms for adaptive targeted web search.

The interaction with the system is maintained by Conversational module [10] which can be serving users or external consumer applications over different communication channels using controlled language [11] which can be wrapped into REST JSON API and/or API expressed in terms of the same language. The search operations can be handled in few following ways.

Normally, the latent Web search operations are executed periodically given configured URLs and patterns for topics selected by user explicitly with obtained results being ranked on the preferences inferred from feedback. In the existing [8,10] agent application described further the URLs and patterns can be also extracted and configured automatically due to background cognitive activity of the agent application profiling user's activities in social networks or on mobile device. If the patterns are matching during the search, they are kept in STM as well as serialized to LTM. If the pattern contains variables, the variable slots are filled and respective match turns into object accumulating relationships to the entities represented by variables. The findings are associated with link to Web page of origin, optional image taken from the page, and time stamp. Then the findings can be used to build user's news feed in user interface or RSS channel. In addition to that, if user has integration with any chat/messenger application or email notification turned on, respective alert is sent to user into chat or to email address.

Further, any search request from user directed via chat, messenger, email or user interface can be searched in memory of the system in few different ways, based on the preferences of the user or depending on availability of results. By default, the search query or pattern specified by user ad hoc is initially searched in existing findings, and returned if found. If no such finding is present the Temporal Graph (Fig. 1.) containing inverse indexes from words to cached pages is used to lookup potentially relevant ones which are then matched using the same Matcher component and semantic extraction of entity attributes applies ad-hoc in case if there are variables in the pattern of query.

In addition, it may be requested to search particular Web site using the same adaptive targeted algorithm and semantic extraction with results delivered to chat interface or messenger or email upon the search completion.

Moreover, the self-reinforcement mechanisms may be used based on feedback given by user in respect to returned findings. For each of the newsfeed items, user may check if an item is fitting the criteria actually. In such case, checked matching results in newsfeed are involved in latent pattern mining process to figure out other patterns that may be potentially serving user's interests. Also, the text features of the summaries generated by Matcher based on on older checked findings are used to compute "personal relevance" of the new findings suggested by user so the items in news feed are ranked ac-cordingly to "personal relevance". That is, the search engine is providing greater personalization based on reinforcement learning.

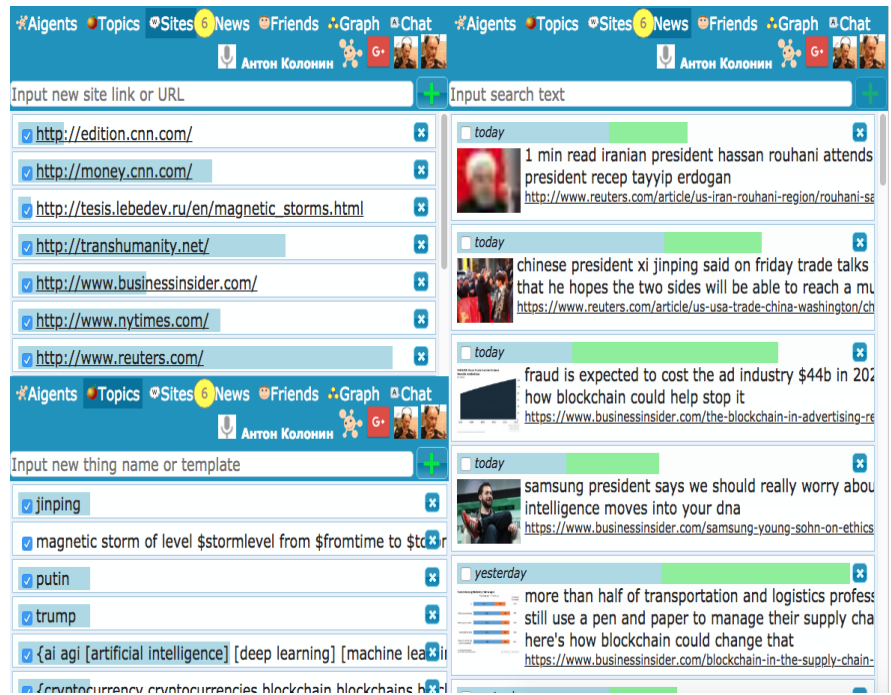

**Fig. 4.** Screenshots of the Web client application enabling user to select starting URL-s for the search and indexing (left on the top) topics and query patterns of the interest for latent search (left on the bottom) and view the news feed (right) with ability to give either negative feedback removing item or providing positive feedback placing check mark on the item. The left screenshot also represents view of the "personal relevance" (left blue horizontal bar width) and "social relevance" (right green horizontal bar width) computed on basis of feedback provided by particular user and trusted social connections if given user, respectively.

If there are more than one user using the system at the same time, and if there are trust connections established by users, the system is also able to provide extra recommendation power based on "social relevance", based on feedback to the findings and features of the findings given by other users.

All of the processes described above are designed to be adaptive to both time and memory constraints of given hosting device or environment used for deployment of the system.

The breadth and depth of search per starting URL and per target topic pattern are bound to time quota computed given entire scope of URLs and topics scheduled for latent search given period of time. In fact that means that breadth and depth of search are bound to amount of computational resources available to system at a time.

The temporal ranges of information indexed for sequential search on time scale are bound to memory resources available as well, so that amount of storage space limits time range of persistent Temporal Graphs while amount of RAM memory affects capacity to keep subgraphs relevant to attention focus in STM. The former means that the larger storage volumes make it possible to perform retrospective searches at more distant past. The latter means the larger amounts of RAM make operations more time-efficient.

## 3 Practical Probation

The practical probation of the system described above is being made in the course of Aigents project available in Java as open source https://github.com/aigents/aigents-java as personal autonomous agent intended for Web search via either locally installed desktop application or shared

Web application running on cloud hosting. Mobile version of the implementing application is now available for Android as open source as well https://github.com/aigents/aigents-android. The Web user interface to demonstration server instance is also available at https://aigents.com/ with screenshots of the interface presented on Fig. 4.

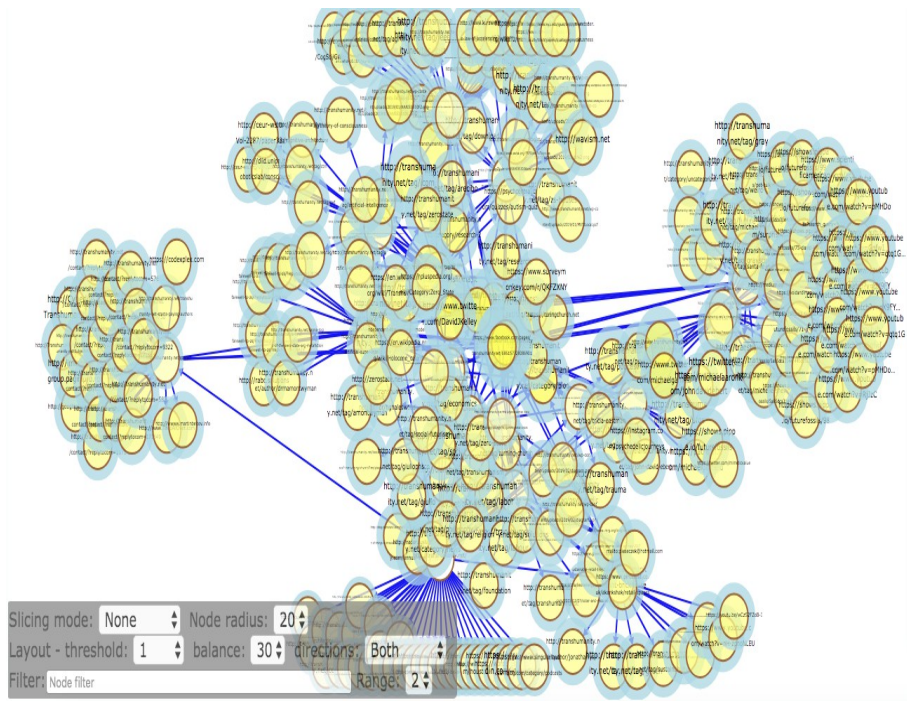

**Fig. 5.** Temporal slice of Web site connectivity graph in range of 2 hops around http://transhumanity.net/ site, which is possible for querying via system API for selected time intervals and sub-graph retrieval parameters such as number of hops, maximum number of nodes and parameters of the graph rendering.

Typical use case of the personal agent is the following. On first interaction, agent extracts user profile in either web search and browsing history on mobile device or history of interactions on social networks on the Web. Based on clustering and pattern mining in that volume of information, initial set of starting URL and topic patterns for latent Web search is created and the search activity is started. Further, the user is able to edit his or her profile – adding, deleting or editing starting URLs and topic patterns. Also user provides feedback on the news feed items coming up on regular basis and submits ad-hock query requests upon the need.

The system is capable to perform well in both mobile single-user and server due to the system adaptability to handle different RAM memory and persistent storage environments, having parameters of search adapted to available resources.

Another side-feature of Temporal Graphs used to maintain and explore indexes is temporal analytics of connectivity dynamics of sites and pages as shown on Fig.5.

With the latter feature, contents of the graphs representing Web site connectivities via links and Web page patterns in terms of text features used on these sites may be explored with graph manipulation operations supported by the API – such as retrieval of subgraphs with selected list of "seed" nodes, selected types of links to follow, limit on number of hops and total amount of nodes and links.

The other practical feature of the system is making it possible to have any custom news feed to be configured as RSS feed available for the public use, like the http://aigents.com/al?rss%20ai RSS channel dedicated to AI, AGI and blockchain technologies is created on base of custom instance of personal agent hosted at https://aigents.com/.

Given the Web and Android applications are being used by almost seven hundreds of users we anticipate that solution would be useful for any users concerned about having high degree of personalization and time-critical updates on the matters of their interests having the privacy and security of the search preserved. It is also expected that this kind of search solution would be demanded by vendors of autonomous agents of AI with need for autonomous search and monitoring of the online information and information extraction from the retrieved sources.

## 4 Conclusion

The presented architecture of the adaptive targeted latent semantic web search with reference implementation suitable for mobile, desktop and cloud server deployments is available as open source distribution.

It applies practical for both personal and corporate use and may be employed by third-parties building AI and robotic solution with need for Web search functions or capabilities to monitor online information.

Our future work will be dedicated for extending adaptability of the described framework and providing support or greater variety of information sources such as social networks and messaging platforms, which is partially supported already.